\documentclass[letterpaper]{article}
\usepackage{iccc}
\usepackage{graphicx}
\usepackage{verse}

\usepackage{times}
\usepackage{helvet}
\usepackage{courier}
\usepackage{comment}
\title{
Creativity as a Human Right: Design Considerations for Computational Creativity Systems}

\author{Alayt Issak\\
College of Arts, Media and Design (CAMD)\\
Northeastern University\\
Boston, MA 02115  USA\\
issak.a@northeastern.edu\\
}
\begin{document} 
\maketitle
\begin{abstract}
\begin{quote}

We investigate creativity that is underlined in the Universal Declaration of Human Rights (UDHR) to present design considerations for Computational Creativity (CC) systems. We find this declaration to describe creativity in salient aspects and bring to light creativity as a Human Right attributed to the Fourth Generation of such rights. This generation of rights attributes CC systems and the evolving nature of interaction with entities of shared intelligence. Our methodology examines five of thirty articles from the UDHR and demonstrates each article with actualizations concluding with design considerations for each. We contribute our findings to ground the relationship between creativity and CC systems. 

\end{quote}
\end{abstract}

\section{Introduction}

This work joins the evolving conversation on the need to model values that computationally creative systems theoretically and practically build upon \cite{saunders_multi-agent-based_2019,kaplan_intrinsically_2006,cassion_humble_2021,seymour_respect_2022}. Creativity is a concept that has received much knowledge and acceptance to generalized understanding, akin to an art piece that has multiple representations \cite{said_metwaly_methodological_2017}. To this value, we examine creativity exhibited in Computational Creativity (CC) as CC warrants an eye that adjoins humanity with entities of shared intelligence. Hence, we present \textit{creativity as a human right} to ground discussions on CC systems where the perspective serves as a bedrock to addressing interactions with these systems. We do so by introducing the Universal Declaration of Human Rights (UDHR) as our foundation of rights, weaving creativity throughout selected articles in consideration of CC systems.

In tangential work, \citeauthor{loi_society_creative_2020}~(\citeyear{loi_society_creative_2020}) present the societal and ethical relevance of CC with AI systems that enable creativity. Noted as ``creativity-enabling AI'' and divided among the categories of natural, social, and internal existential creativity, the authors elicit one outcome of such to be ``CC with resilience''. \citeauthor{smith_neu_gsnd_2017}~(\citeyear{smith_neu_gsnd_2017}) also presents a stance for CC systems for social change, i.e., how we consider computationally creative projects in relationship to values associated with social justice and how our own biases as humans are threaded into these systems. 

In summary, we put forward the need to look at creativity as a human right to further the field of Computational Creativity. This approach secures upstream impacts on grounding creativity from the evolving perspective that is the nature of CC. We present our contribution to design considerations for CC systems that address open questions such as the formalization of creativity with emerging creative agents.

\section{Background}

\subsection{Generations of Human Rights}

Human Rights evolve through a series of generations in their complication and jurisdiction with the recognition and implementation of innate needs. In doing so, they currently sit within four generations. According to a review on \textit{The Generations of ``Human’s Rights''} \cite{cornescu_generations_2009}, their recognition is situated from a shift in antiquity, where the balance of individual rights was in favor of the state, to the medieval period where the monarchy held absolute power. 

As such, the First Generation of rights emerged from a return to jusnaturalist conceptions of what these rights ought to be. This generation, which fell in the age of Western Enlightenment, highlights civil and political rights that fall under subjective rights for inherent human rights. These entail the freedom of opinion, expression, personal ownership, political security and citizen participation in power \cite[p.~4]{cornescu_generations_2009}.

Following the First Generation arose socio-economic and cultural rights to which work, freedom of association, social rights such as social security, medical services, pensions, and education were taken into consideration. This Second Generation of rights highlights the shift from the personal autonomy of the First Generation in ``free status'' to a ``social status'' upon the social situatedness of citizens \cite[p.~5] {cornescu_generations_2009}.

Following the association of the Second Generation, the Third Generation emerged where rights moved towards the identification of ``solidarity'', where they can no longer be implemented on their own, but in a collective effort. This feature contradicted earlier generations for its compromise of individual and social needs. Rights in this generation collectively entail those to self-determination, peace, development, and humanitarian assistance. A key feature of this generation entails those focused on ``the rights of future generations'' that take into account ``the needs of the present without compromising the ability of future generations to meet theirs'' \cite[p.~6]{cornescu_generations_2009}. 

With the Fourth Generation arose the current conversation around what should be implemented, in the application of ``what is right'' and its contradiction with what is known. Questions in this generation raise the legal, ethical, moral, and religious issues that emerge with the implementation of unseen innovations such as creative augmentation, genetic manipulation, artificial life, and their placement among the previous generation of values \cite{risse_fourth_2021}. The generations of these rights assess humanity from a speculative perspective that is forward-looking while referring to the past of what determines these novelties to be human rights in the first place. 

Epistemic rights, ways of knowing in exercising human intelligence, are proposed in this generation. This set of rights entails humanity’s relationship with entities of similar or larger intelligence such as CC systems within AI and automation \cite[p.~362]{risse_fourth_2021}. Most recently, a first-of-its-kind workshop was held at the 2023 ACM Creativity \& Cognition conference to unravel participant perception of the interaction between AI and the Arts \cite{makayla_aixartist_2023}, addressing CC systems' placement in this generation of rights. 

\subsection{Creativity}

\begin{figure}[h!]
    \centering
  \includegraphics[width=0.9\columnwidth]{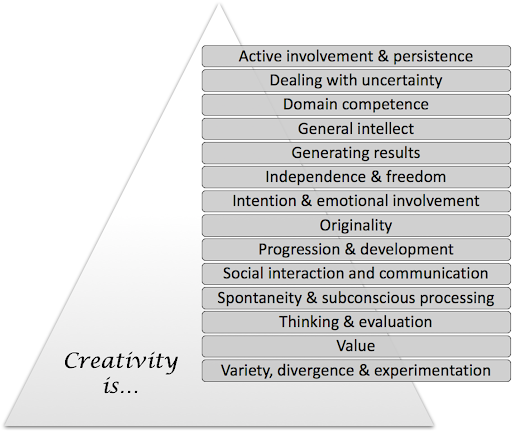}
  \caption{Fourteen key themes and factors of creativity \cite[p.~18] {jordanous_components_2016}. Referred to in numerical order.}
  \label{fig:diagram}
\end{figure}

We refer to the survey work - \textit{Modelling Creativity: Identifying Key Components through a Corpus-Based Approach} - as a detailed approach to our working definition of creativity \cite{jordanous_components_2016}. We utilize the fourteen themes of creativity identified in the paper (Figure \ref{fig:diagram}) to present creativity as a human right. In doing so, we elicit the aforementioned themes, aligning creativity with the UDHR and integrating how the declaration touches upon creativity to explicitly address CC systems. 

\section{Methodology}

Of the 30 articles present in the UDHR, it is the fourteen themes from the survey work that lead to the UDHR's alignment with creativity. We present this insight as we examine five articles of the UDHR, namely Articles III, XVIII, XIX, XXII, and XXVII. These articles were decided upon reading through the declaration for relation and relevance to creativity. Our articles also investigate Epistemic Rights of the Fourth Generation in the overlap with Articles XVIII, XIX, XXII, and XXVII \cite[p.~359-363]{risse_fourth_2021}. We also refer to the International Covenant on Civil and Political Rights (ICCPR), to add upon a similar document to the UDHR, with shared positions among Articles I, X, XVIII, and XIX \cite{ohchr_civil_and_political_1966}.

\section{The Universal Declaration of Human Rights (UDHR)}

In this section, we detail the psychological underpinnings of creativity and its social situatedness within the UDHR. We outline Articles III, XVIII, XIX, XXII, and XXVII as our framework for analysis, associating each article to its relationship with CC systems respectively. 

\subsection{Article III}

Article III begins with a protection for the security of person, and human mental suffering with respect to the law. However, in this sense, can also be alluded to protect the mental needs of individuals. That is, an act of creativity as mental ideation touched upon in the precedence given to mental thought. Article 10 of the International Convention on Civil and Political Rights (ICCPR) also details the inherent dignity of the human person for those deprived of their liberty to be treated with humanity.

Creativity summons an eye into the notions at which mental health is preserved in this action. James Kaufman~(\citeyear{kaufman_aeon_2023}), a psychologist of creativity and Professor of Educational Psychology at the University of Connecticut, notes creativity as a way of life that lives in the interior as opposed to the Big-C creativity that permeates the act. This refers to the third theme of creativity in domain competence and the fifth theme of generating results. Mental suffering that is associated with creativity by the need to produce, whether within or outside, brings into light what one must protect from oneself, which is the base of the warrant by Article III in maintaining creativity within one’s life and the second theme of dealing with uncertainty.

Creativity to mental suffering is also often associated with acts of resistance and a response to suffering. These acts entail repressed humanity that elicits responses from an expression of those inheritances, such as occurrences in the restriction of autonomy and dignity. Such acts that have deprived human beings historically arise in the atrocities of slavery, which in the Americas, from 1619 to abolishment in 1865, conjured horrifying acts on Native Africans kidnapped from their home countries and brought to North and South America for political and economic incentives \cite{laveist_400_2019}. 

In doing so, creativity emerges as an eye to abstract suffering of mental capacities. It creates a mode of understanding for ``suffering rooted in suffering''. It is a lens through which one sees the world that is warranted in expression and communication. Novel output and that which is deemed socially acceptable by recipients, creativity is the lens of suffering through which endurance holds its ground and humanity expresses its deepest conjuring. 

In this observation of human denial, creativity warrants the expression for such a sense and a human right to which those who reside in the restriction of their humanity, bear to express their means, emotions, and suffering. As such, we posit CC systems to be respondents and inclusive of this expressive embodiment as a means to elicit the interactions of such framings. Analysis of embodiment in CC research has previously examined such approaches for the ``perception of creativity'' \cite{moruzzi_artificial_2022,guckelsberger_embodiment_2021}, yet leaving open questions on the embodiment of CC systems themselves with creativity as a human artifact. 

\subsection{Article XVIII}

Article XVIII of the UDHR and Article XVIII of the ICCPR state one's freedom of thought and conscience with respect to religion, where religion elicits ideas that amalgamate beliefs. Rooted in spirituality, it is what defines the connection to the outside world, where the arts, most often seen as a spiritual practice, are categorized as non-religious spirituality \cite{gautam_spirituality_2017}. One finds art in a spiritual path, a path to reconciliation, a place of refuge. 

The arts are an aspect in which the freedom to creativity is positioned in spirituality \cite{coleman_creativity_1998} and the eleventh key theme of subconscious processing. Expressions that propagate the spiritual element of art can be traced to Carl Jung’s denomination of art states as the natural and primary language of the psyche \cite[p.~5]{gautam_spirituality_2017}. Abstract theorist Wassily Kandinsky, in the book \textit{Concerning the Spiritual in Art}, highlights art as the spiritual revolution that manifests through the posits of color and composition \cite{kandinsky_concerning_2012}, and to those partaking within the act, the furthering of transience and transcendence that elicits the complexity of human consciousness, a feature detailed by Alex Grey in \textit{The Mission of Art} \cite{grey_mission_2001}. 

Spiritual exposition to the arts is extended as an innate part which in itself is a sub-category, a secular release, that is embodied in relation to the spirituality of one’s being. Enacted within religion to the human rights observed within these articles, it is this element that is salient in the article. Hence, to abstract beyond the presence of religion, it is the dual element of spirituality that exists, that to its paradoxical nature, also acts as the primary nature in which religion expresses its transience \cite{mitias_creativity_1985}. Religious iconography, such as Islamic Calligraphy, that elicits the meanings of worship and transcriptions of scripture are examples of such exhibitions \cite{nasr_islamic_1987}. To many of its constituents, the creative act of expressing religion is a capture of the psyche and a release of the spiritual element for binding the presence of creativity, which exists in spirituality and art. It is the process that is central by which one dwells with situations of uncertainty, instability, uniqueness, and value conflict \cite[p. ~50] {schon_reflective_1983}. In such an approach, we argue CC systems to be canals that carry forth these processes for the situations they respond to and for those that exist within the subconscious creative act. 

\subsection{Article XIX}

Article XIX states the freedom of opinion and expression in the UDHR. This is equally stated as the elicitation of opinions without interference in the parallel article (XIX) of the ICCPR. Although framed in the lens through which one accesses information with the freedom to accept and interpret information, it is a need that imparts the expression linked to the medium of the arts. The core of the law which states the freedom of expression, towards one's ideas, implements creativity as an expression of oneself with outlets, the primary example being art. 

A notable example of this is the Post-modernist artists, to which expression of reality and truth was presented in flux. This encounter leads to the fourteenth theme of divergence and experimentation. Each art movement, as it drifted from Impressionism to Dada and Surrealism, carried with it expression and a worldview to what it sees. The Impressionists sought to present what the eye literally saw, an act that was not so confronted by society to the goers of art exhibits, yet the Dadaists in response to WWI saw the calamity of the world, presenting information about reality with conceptual art on the craft and knowledge of nonsense \cite{bonnett_art_1992}. From the literal ``nonsense'' in Dada to Expressionism’s abstract paintings, it is information that is distilled through the arts at which creativity, for social interaction with the world, bounds the upheaval to which these ideas came to be. 

This resultant underpinning of reality with the expression of creativity in interaction and involvement is presented as a key element for CC systems to embody within this article of rights. Levels of interaction bound upon levels of autonomy foundationally distinguish the eliciting of such expression \cite{daniele_ai_2019}, whereby expressive mechanisms such as the sublime \cite{crowther_sublime_2016} would adhere CC systems that not only allude to individual expression but also to the formalization of creativity that is intertwined with creative agents.  

\subsection{Article XXII}

Taking a societal approach, Article XXII presents the cultural rights to which human rights are bestowed - ``the right to social security and realization of economic, social and cultural rights indispensable to his dignity and the free development of this personality.'' This is detailed as self-determination in Article I of the ICCPR. The latter half, which is the free development of personality, is intertwined with the cultural regulations of one's inner being. Culture instills creativity within the lens of what is acceptable and what is not. 

Within a larger framework, this article lies in one's ability to present oneself regardless of the culture one inhabits. It is the desire for self-expression and the formalized conception of self-actualization \cite{ryder_role_1987}. Tied with the ninth theme of progression and development, the creative act is linked with one's inner being, towards the self-discovery and the process which acquires emotional, spiritual, and intellectual enhancements to the individual \cite[p. ~6]{manheim_relationship_1998}. It is a process in which the actualization occurs, with the arts as a mechanism of outward expression. 

As such, self-actualization in the arts exhibits itself in symbolisms and incumbently abstractions that garner and question meaning. Drawing literature from \textit{Artistic and Perceptual Awareness}, the process elicits four stages of creative expression through awareness, focus, the working process, and the art product respectively \cite{linderman_developing_1979}. Thus, to the sensitivity and development of creativity, the liberty creativity offers to the changing nature of the inner self is one we argue CC systems integrate as propagators of actualization and the discovery that lies underneath. 

\begin{figure}[h!]
    \centering
  \includegraphics[width=0.6\columnwidth]{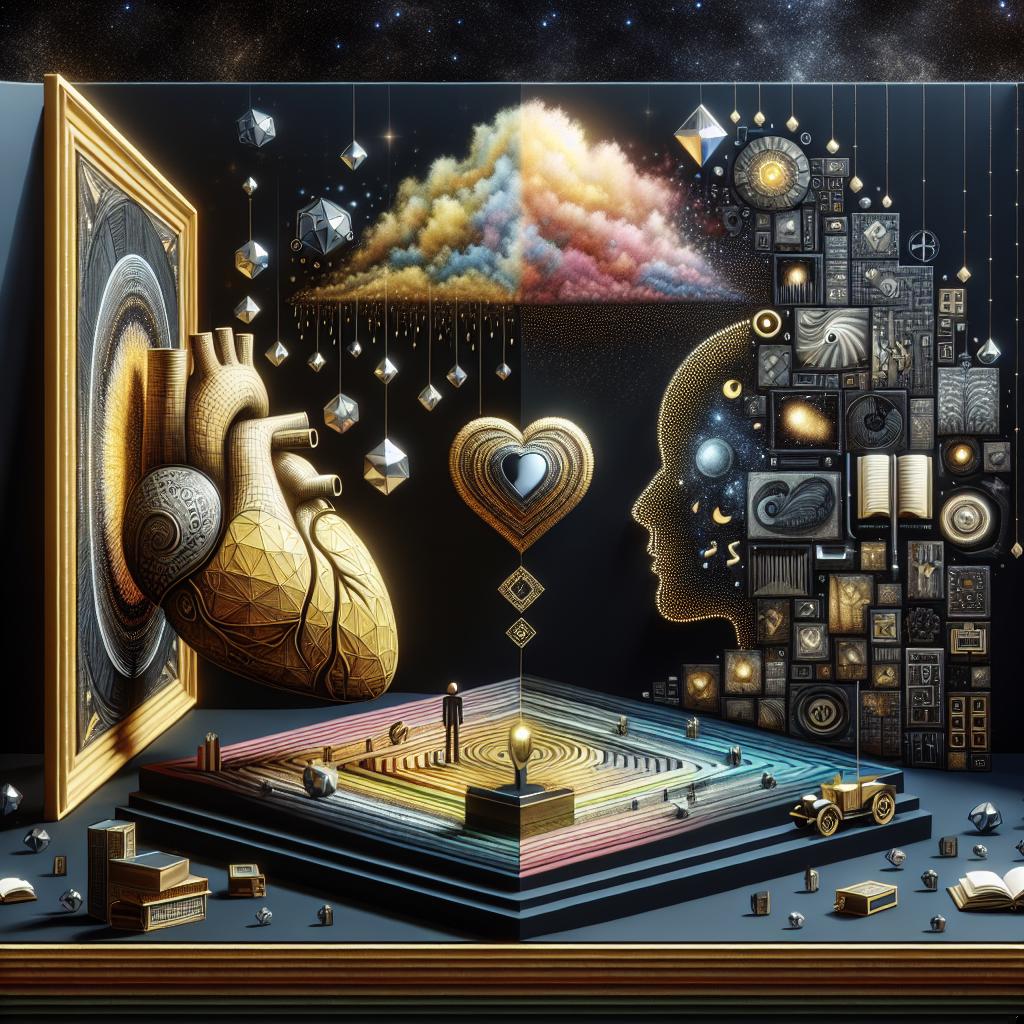}
  \caption{Image generated by the author using T2I model DALL-E 3 \cite{betker_improving_dalle3}. CC abstract generation in the four stages of creative expression - Ideation for \textit{awareness}, \textit{focus} on prompt design, \textit{working process} for prompt engineering, and the \textit{art product} for eliciting an output. Prompt based on abstractions in meaning and questioning.}
  \label{fig:hierarchy}
\end{figure}

Figure \ref{fig:hierarchy} solicits a CC system, text-to-image (T2I) model, with what would have been an internet search of ``abstract imagery to find the process of discovery''. Upon iteration, this is realized into an image that conveys the process of questioning. We present this image to position an interaction with CC systems alluding to the Fourth Generation of Human Rights.

\subsection{Article XXVII}

Upon exaltations of the loose notions of art, comes the pinnacle at which art is directly represented in the UDHR. In its utility, the article has been called upon to protect artists against legal action for artistic freedom \cite{hrc_art_2013}. The article states:

\begin{itemize}
    \item ``Everyone has the right freely to participate in the cultural life of the community, to enjoy the arts and to share in scientific advancement and its benefits''
    \item ``Everyone has the right to the protection of the moral and material interests resulting from any scientific, literary or artistic production of which he is the author''
\end{itemize} 

We divide this clause into two sections and take a deep dive into the lens of creativity that it deployed to the first section, which awards the right to the recipient (``enjoy the arts''), and the second section which attributes the right to the producer (``moral and material interests''). The article warrants on an equal pedestal the freedom to scientific, literary, and artistic pursuits, presenting the arts viewership as a role that is of equal governance to one's human needs. Elaborating on the division is the process of creativity as an act that is involved and immersed in the intermingling, hence incorporating the first theme of active and emotional involvement of creativity. 

\begin{figure}[ht]
    \centering
  \includegraphics[width=0.6\columnwidth]{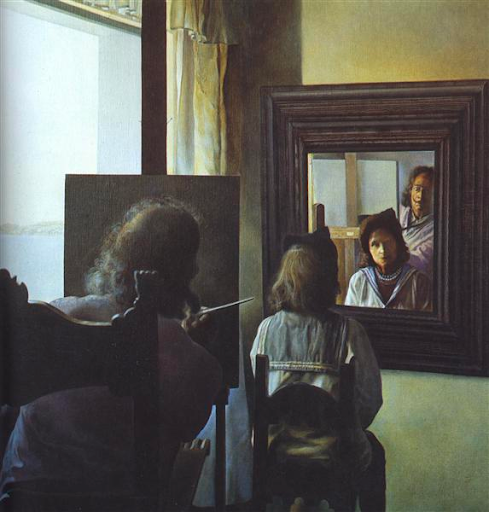}
  \caption{Dalí from the Back Painting Gala from the Back Eternalized by Six Virtual Corneas Provisionally Reflected in Six Real Mirrors, 1972 (Dalí, 1972) \cite{dali_mirrors_1972}.}
  \label{fig:dali}
\end{figure}

Depicted by Surrealist Salvador Dalí in Figure \ref{fig:dali}, creativity is the process, witness, and practice through which the artist is the lens seeing the image. Self-portrait and detailing the viewings of which the artist in the image is visualized conceptualizing about and painting in the act. We present this surrealist image as it presents the two sections collectively and details art, to which the duality of creativity is inherently recognized. Eliciting the notion through the self-portrait of an artist painting and witnessing the act, it is the very act at which creativity exists in the duality of the process and interaction -- a duality we instill in CC systems through their immersion with creativity. 

\section{Conclusion}

In this paper, we examine the UDHR to decipher creativity through facets of humanity that align with CC systems. Attributed within five articles of the declaration, alongside the key themes of creativity, we highlight how creativity lies in the act of mental need, freedom of thought and expression, development of personality, self-actualization, and the going and undergoing of artistic expression. In doing so, creativity is the process that upholds inner humanity and for that grounds itself as a human right for the foundation of CC systems. We present our findings regarding the Epistemic Rights in the Fourth Generation of Human Rights (interactions with CC systems) and ongoing conversations on design considerations for CC systems. 

\section{Acknowledgments}
I would like to thank dan brown, Erica Kleinman, Casper Harteveld, H. C. Robinson (Northeastern Law), and Dietmar Offenhuber for the insightful discussions and valuable feedback. 

\bibliographystyle{iccc}
\bibliography{iccc}

\end{document}